# Semimetallicity and Negative Differential Resistance from Hybrid Halide Perovskite Nanowires

Muhammad Ejaz Khan, Juho Lee, Seongjae Byeon, and Yong-Hoon Kim*

School of Electrical Engineering and Graduate School of EEWS, Korea Advanced Institute of Science and Technology, 291 Daehak-ro, Yuseong-gu, Daejeon 305-701, Korea.

In the rapidly progressing field of organometal halide perovskites, the dimensional reduction could open up new opportunities for device applications. Herein, taking the recently synthesized trimethylsulfonium lead triiodide $(CH_3)_3SPbI_3$ perovskite as a representative example, we carry out first-principles calculations and study the nanostructuring and device application of halide perovskite nanowires. We find that the one-dimensional (1D) $(CH_3)_3SPbI_3$ structure is structurally stable, and the electronic structures of higher-dimensional forms are robustly determined at the 1D level. Remarkably, due to the face-sharing $[PbI_6]$ octahedral atomic structure, the organic ligand-removed 1D $PbI_3$ frameworks are also found to be stable. Moreover, the $PbI_3$ columns avoid the Peierls distortion and assume a semimetallic character, contradicting the conventional assumption of semiconducting metal-halogen inorganic frameworks. Adopting the bundled nanowire junctions consisting of $(CH_3)_3SPbI_3$ channels with sub-5 nm dimensions sandwiched between $PbI_3$ electrodes, we finally obtain high current densities and large room-temperature negative differential resistance (NDR). It will be emphasized that the NDR originates from the combination of the near-Ohmic character of $TMSPbI_3$-$PbI_3$ contacts and a novel NDR mechanism that involves the quantum-mechanical hybridization between channel and electrode states. Our work demonstrates the great potential of low-dimensional hybrid perovskites toward advanced electronic devices beyond actively-pursued photonic applications.

# 1. Introduction

Organic-inorganic hybrid halide perovskites have recently emerged as prominent candidates for optoelectronic applications due to their high light absorption coefficient, long charge carrier diffusion length, intense photoluminescence, slow non-radiative charge recombination rate, and defect tolerant nature.[1-5] Together with their low cost and facile fabrication process, these excellent material properties allowed hybrid perovskites to show enormous potential in optoelectronic device applications such as solar cells, light-emitting diodes, lasers, and photodetectors.

Particularly, as a method to achieve novel material properties and optimized device performance, organic-inorganic hybrid perovskites have been recently made into atomically thin two-dimensional (2D) sheets and one-dimensional (1D) nanowire structures.[6, 7, 8] These low-dimensional halide perovskites with strong quantum confinement and spatial isolation effects have demonstrated great promise for the tuning of photoluminescence peaks and the improvement of device performances.[9, 10] In fact, low-dimensional 2D and 1D forms of halide perovskites were an active research target about two decades ago[11-13] and are recently regaining interest in view of their optoelectronic device applications. Overall, compared with their photonic counterparts, research on the electronic devices based on both three-dimensional (3D) and low-dimensional hybrid halide perovskites has been scarce.

In this work, we explore the novel material and device properties of low-dimensional hybrid halide perovskites in view of realizing advanced electronic devices, and predict the emergence of semimetallic behavior from the inorganic framework of 1D halide perovskites and from their heterojunctions excellent negative differential resistance (NDR) characteristics. As a specific case, we consider recently synthesized trimethylsulfonium (TMS) lead triiodide $(CH_3)_3SPbI_3$ (TMSPbI$_3$) perovskite, which showed an 1D-based crystal structure and good ambient stability.[14, 15] In terms of the atomic structure, TMSPbI$_3$ consists of hexagonal stacks of 1D columns made of face-sharing [PbI$_6$] octahedra, and individual anionic 1D PbI$_3^-$ columns are surrounded and isolated by the organic $(CH_3)_3S^+$ cations. Despite the hint of possibly deriving isolated 2D and 1D nanostructures from the parental TMSPbI$_3$, so far the structural, electronic, optical properties of 1D TMSPbI$_3$ and its 2D network have not been considered yet. Carrying out density functional theory (DFT) calculations, we confirm that 2D monolayer and 1D nanowire structures derived from 3D TMSPbI$_3$ are dynamically stable, and the electronic structures of 3D and 2D TMSPbI$_3$ analogs are essentially determined by that of the 1D component. Most intriguingly, we find that the inorganic 1D PbI$_3$ framework remains structurally stable even after organic ligands are stripped off, and furthermore it exhibits a semimetallic character. Having structurally stable and yet electronically distinct nanowire components out of the bulk TMSPbI$_3$, we finally devise bundled nanowire heterojunctions in which semiconducting TMSPbI$_3$ channels with sub-5 nm dimensions are sandwiched between semimetallic PbI$_3$ electrodes. Carrying out DFT-based non-equilibrium Green's function (NEGF) calculations, we then obtain strong NDR properties characterized by high current density (in the range of ~ 900 kA.cm$^{-2}$) and peak-to-valley current ratios (PVRs, up to ~17) at low-bias regimes (< 0.5 V). It will be emphasized that the predicted NDR results from the combination of a novel NDR mechanism based on quantum-mechanical hybridization of channel electronic states and the unique near-Ohmic character of PbI$_3$-TMSPbI$_3$ contacts.[16]

# 2. Results and Discussion

## 2.1 *Structure and stability of TMSPbI$_3$ and PbI$_3$ nanowires*

In Figure 1a-c, we show the atomic structures of 3D bulk, 2D nanosheet, 1D nanowire TMSPbI$_3$ perovskites, respectively, optimized within the Perdew-Burke-Ernzenhof parameterization of generalized gradient approximation revised for solids (PBEsol).[17] The inorganic PbI$_3$ nanowire structure prepared by removing ligands is additionally presented in Figure 1d, and we will later show that it is also a dynamically stable structure. The experimentally synthesized TMSPbI$_3$ 3D crystal structure has hexagonal symmetry in the space group *P6$_3$mc* (no. 186) with a 1D network of PbI$_3$ containing distorted face-sharing octahedra along the *c* axis, and the 1D PbI$_3$ columns are surrounded by two trimethylsulfonium $(CH_3)_3S^+$ ligands per unit cell (UC).[14, 15] In the 1D TMSPbI$_3$ nanowire, due to the removal of steric hindrance imposed by neighboring 1D nanowires, the ligands are rotated and shifted slightly away from the positions in 3D and 2D structures and results in a noticeable symmetry breaking with respect to the axial planes (Figure 1c). On the other hand, after removing the ligands, we observe that the 1D PbI$_3$ recovers the highly symmetric structure (Figure 1d; see also Supporting Information Figure S1 for a detailed analysis of 1D PbI$_3$ crystal structure).

To explore the feasibility of obtaining isolated 2D and 1D nanostructures from the experimentally-synthesized 3D form, we further calculated inter-layer and inter-column binding or exfoliation energies for the former and the latter cases, respectively. We obtained the inter-layer binding energy of 17.7 meV·atom$^{-1}$ for the 2D layer and the inter-column binding energy of 44.8 meV·atom$^{-1}$ for the 1D nanowire. Our calculated interaction strength for 2D and 1D nanostructures lies within the range that permits the mechanical exfoliation to synthesize free-standing nanostructures in experiments (≲150 meV/atom).[18] These inter-layer and inter-column binding energy values were compared with DFT-D3[19] results, and we confirmed that they remain within the experimentally accessible range for exfoliation (For the details, see Supporting Information Table S1).

Finally, in addition to the 2D and 1D nanostructuring, we also considered the feasibility of obtaining stable inorganic $PbI_3$ columns by removing organic ligands from $TMSPbI_3$. Calculating the ligand binding energy with reference to the trimethylsulfonium iodide ($(CH_3)_3SI$), which was used in experiments as the precursor to synthesize $TMSPbI_3$,[14, 15] we obtained −1.1 eV·formula-unit$^{-1}$. While the magnitude is in line with the unique role of the sulfur-based ligands to stabilize the octahedral face-sharing 1D $PbI_3$ frameworks during the synthesis of $TMSPbI_3$, it is comparable to the widely exploited gold-thiolate surface binding energies.[20-22] So, we expect that it will be possible to selectively remove $(CH_3)_3S^+$ ligands from $TMSPbI_3$ through chemical cleaning processes such as the controlled ozone treatment.[23]

Having confirmed the feasibility of preparing isolated 2D sheet and 1D nanowire $TMSPbI_3$ forms, we further explored their dynamic stabilities by calculating their phonon spectra (Figure 2; see Supporting Information Figure S2 for the 3D case). We find that imaginary phonon modes are absent in the phonon band dispersion curves, which confirms the stability of 2D and 1D nanostructures. Regarding the role of $(CH_3)_3S^+$ ligands, we observe that unstable phonon modes emerge when one of the two ligands in the 1D UC is removed (Supporting Information Figure S3) and conclude that the isometric distribution of ligands around the inorganic 1D $PbI_3$ column is required to structurally stabilize $TMSPbI_3$. Most interestingly, however, if all the $(CH_3)_3S^+$ ligands are removed, the inorganic 1D $PbI_3$ framework becomes again a dynamically stable structure as shown in Figure 2c. The mechanisms of the structural stabilization of $PbI_3$ will be discussed below together with the analysis of its electronic structure.

## 2.2 Electronic and optical properties of $TMSPbI_3$ nanowires

In Figure 3a and 3b, we respectively show the calculated electronic band structures and optical absorption spectra of 2D and 1D $TMSPbI_3$ perovskites as well as the ligand-removed 1D $PbI_3$ (see also Supporting Information Figure S4 for the bulk $TMSPbI_3$ data). Here, to correct the well-known bandgap underestimation within the local and semi-local DFT exchange-correlation functionals, we employed the Heyd-Scuseria-Ernzerhof (HSE) hybrid functional.[24] In addition, due to the presence of heavy Pb atoms, we included spin-orbit coupling (SOC) effects, which resulted in the splitting of conduction band bottom energy bands and the resulting reduction of band gaps compared to the non-SOC values (see Supporting Information Table S2 and Figure S5 for the detailed comparisons between the results obtained without and with SOC). Then, for the bulk $TMSPbI_3$ case, we obtained the bandgap of 3.25 eV, which is in good agreement with the experimental optical bandgap of 3.15 eV.[14, 15] Experimentally, this bandgap value was extracted by assuming an excitonic absorption, or excluding the peak right above the absorption onset, and our calculation of optical spectra that neglects local field effects provides the direct confirmation of the assumption (see Supporting Information Figure 4Sb). The subband gap excitonic peak in the measured optical spectra is rather prominent.[14, 15] While we do not attempt to reproduce this excitonic absorption peak by resorting to the many-body approach or time-dependent DFT formalism,[25, 26] it clearly indicates the large quantum confinement effect even in the bulk limit or the electronically decoupled nature of individual 1D $TMSPbI_3$ columns.

The comparison of 1D and 2D $TMSPbI_3$ band structures (Figure 3a and Table 1) and optical spectra (Figure 3b) with those of their 3D counterpart (Supporting Information Figure S4) show the detailed nature of quantum confinement effects. Specifically, we find that the electronic structures of the 3D and 2D $TMSPbI_3$ are essentially identical to that of the 1D analog. The intrinsically 1D nature of $TMSPbI_3$ as well as the above-presented stability of its low-dimensional forms can be understood in terms of the connected face-sharing octahedral structure of the $PbI_3$ framework and the absence of physical sharing between these 1D $PbI_3$ columns. The calculated electronic band structures of the $TMSPbI_3$-derived 3D and 2D structures indicate that they are slightly indirect band gap semiconductors, and that the difference between indirect and direct gaps reduces with the dimensionality, leading to the direct band gap of 3.33 eV in the 1D $TMSPbI_3$ case. Analyzing the projected density of states (DOS) shown in Figure 3a (see also Supporting Information Figure S4a), we find that band edges in bulk and low-dimensional analogs are mainly contributed by the core inorganic region $PbI_3$, whereas the electronic states related to the organic ligands $(CH_3)_3S^+$ are located at the energy ranges away from band edges.

The dissemination of integrated charge density around band edge regions in the 1D case further shows that the valence band maximum (VBM) primarily originates from I $5p$ orbitals, while the conduction band minimum (CBM) is produced from Pb $6p$ and I $5p$ orbitals (Figure 3c; see Supporting Information Figure S6 for more detailed exposition of VBM and CBM electronic wavefunctions at different Brillouin zone points and Figure S7 for the orbital contributions to the DOS of these band edges). Unlike in the 3D and 2D counterparts, we find that the CBM of 1D case is partly contributed by $(CH_3)_3S^+$ ligand states.

We note that this feature is unique to the 1D case, and is in line with the above-discussed distortion of inorganic core framework (Figure 1c) that results in the stronger hybridization between organic ligand and inorganic core states. The calculated optical absorption peaks in 3D (Supporting Information Figure 4Sb) and 2D structures (Figure 3b) exhibit relatively stronger intensity along the octahedral face-sharing TMSPbI$_3$ column axial direction ($\varepsilon_2^{\parallel}$) than along the weakly-connected normal direction ($\varepsilon_2^{\perp}$), and particularly the former is in good agreement with the experimental spectrum shape.

### 2.3 Electronic and optical properties of TMSPbI3 nanowires
*Semimetallicity from PbI3 inorganic frameworks*

We now consider the inorganic 1D PbI$_3$ case, which was earlier found to be dynamically stable even after (CH$_3$)$_3$S$^+$ ligands are removed. Strikingly, we find that it exhibits semimetallic characteristics with a linear dispersion at the Fermi level ($E_F$), which implies that the Peierls distortion is absent in 1D PbI$_3$. To confirm the stability of the semimetallicity in 1D PbI$_3$, we stretched and compressed the inorganic octahedral framework along the *c* axis and observed that the semimetallicity is robustly preserved (Supporting Information Figure S8). On the other hand, considering the 1D PbBr$_3$ column derived from (CH$_3$)$_3$SPbBr$_3$, which has been also synthesized recently,[15] we find that the structure becomes dynamically unstable and goes through a Peierls distortion (Supporting Information Figure S9). To further analhyze the origin of the absence and presence of Peierls distortion in the inorganic PbI$_3$ and PbBr$_3$ columns, respectively, we analyzed the local DOS of PbI$_3$, PbBr$_3$ and symmetry-enforced (unstable) PbBr$_3$ structures. The comparison shows that, while the I 5*p* lone-pair orbitals spatially interact with each other in the PbI$_3$ case, such neighbor interactions do not exist for the Br 4*p* lone-pair orbitals in the symmetry-enforced 1D PbBr$_3$ counterpart. Namely, due to the large atomic size of I, the circumferential I-I interactions will prevent the distortion of [PbI$_6$] octahedral cages in the PbI$_3$ case or suppress the Peierls distortion. Apparently, with the smaller atomic size of Br, the PbBr$_3$ counterpart lacks such a mechanism and undergoes a Peierls distortion or distortions of the [PbBr$_6$] octahedral cages. This could become a useful guideline for the future research on low-dimensional halide perovskite inorganic frameworks.

Finally, in terms of the optical spectra presented in Figure 3b, we note that PbI$_3$ exhibits a plasmonic absorption peak at around 1.6 eV. This nanowire plasmon character suggests the potential of metallic 1D PbI$_3$ for novel applications that were not previously considered for hybrid halide perovskite materials alone.[27, 28] For example, the incorporation of plasmonic nanoparticles in a perovskite film was previously shown to improve the photovoltaic efficiency,[29, 30] and our finding implies that such an effect might be achieved with the interfaces purely based on hybrid perovskite-derived materials. While such studies appear promising, we now introduce a completely new research direction for hybrid halide perovskites, that is, the negative differential resistance effect.

### 2.4 Ultrahigh NDR characteristics from halide perovskite nanowire junctions

Given the presence of novel semiconducting TMSPbI$_3$ and semimetallic 1D-PbI$_3$ nanowires, we carried out DFT-based NEGF calculations and studied the bias-dependent quantum transport processes in bundled nanowire junctions. The transport device model was derived directly from the 3D TMSPbI$_3$ and is composed of semiconducting TMSPbI$_3$ channels sandwiched by metallic PbI$_3$ electrodes as shown in Figure 1e (see Supporting Information Figure S10 for the details). The current-voltage (*I-V*) characteristics obtained for the three to five UC (3UC - 5UC) TMSPbI$_3$ channels are shown in Figure 4a. We observe that, as summarized in Table 1, excellent NDR performances characterized by very high peak current density (reaching ~ 4750 kA·cm$^{-2}$ for the 3UC case) and PVRs as large as a few tens (reaching ~ 17.4 for the 5UC case) are achieved with sub-5 nm long channels in low-bias voltage regimes (< 0.5 V). Moreover, these excellent NDR characteristics were obtained at room temperature, and they were robustly preserved down to the zero temperature limit (see Supporting Information Figure S11).

To understand the mechanisms behind the strong NDR signals, we first show in Figure 4b the zero-bias $E_F$ transmission eigenstates[31] propagating from the left PbI$_3$ electrode. We observe an overall very strong penetration of metallic PbI$_3$ states into the semiconducting TMSPbI$_3$ channel region, which indicates that our device operates based on quantum tunneling. The penetration length of PbI$_3$ states is over two UC length, which explains the exceptionally high current density level as well as rapid initial increase of currents in the 3UC channel case.

Next, to explain the bias voltage $V_b$ dependence of tunneling currents, we show in Figure 4c the development of transmission spectra for the 5UC case with the increasing $V_b$ (see Supporting Information Figure S12 where the transmission spectra for the 3UC, 4UC, and 5UC cases are compared). In Figure 4d, we also show the corresponding molecular projected Hamiltonian (MPH) eigenstates[31, 32] that contribute most strongly to quantum tunneling. We note that until $V_b$ = 0.25 V (the NDR peak point) the MPH eigenstates (−0.091 eV and −0.043 eV for the $V_b$ = 0.15 V and $V_b$ = 0.25 V cases, respectively) are strongly delocalized through the channel region. As long as the delocalized MPH eigenstates or large transmission functions are maintained, the widening of bias window will result in current enhancements. However, upon further increasing the bias voltage to $V_b$ = 0.46 V (the NDR valley point), we observe that the spatial connectivity of MPH eigenstates with the right terminal region is abruptly broken, indicating the suppression of transmissions or reduction of currents.

### 2.5 *The quantum-hybridization NDR mechanism*

We finally analyze the nonlinear *I-V* characteristics in more detail and show that it is a manifestation of a novel NDR mechanism. In Figure 5a-d, we present the development of projected local DOS across a PbI$_3$-5UC TMSPbI$_3$-PbI$_3$ junction with increasing $V_b$. Remind that in the resonant tunneling diode and tunnel diode, which have the largest parallel with our device in that the device characteristics have quantum mechanical origins, the NDR signals arises from the (intraband) tunneling across two barriers in the former case (Figure 5e) and the band-to-band (interband) tunneling in the latter (Figure 5f). However, our perovskite nanowire junctions involve neither two barriers nor the band-to-band tunneling. Instead, the NDR signal in our case is produced by the quantum-mechanical hybridization between left and right electrode states in the low-bias regime ($V_b$ = 0.15 V and 0.25 V) and its abrupt disruption in the high-bias regime ($V_b$ = 0.46 V). Note the strong presence of (two) electrode Fermi-level states penetrating into the channel at finite bias,[33] which are symmetrically tilted until $V_b$ = 0.3 V and then abruptly broken into an asymmetric form at $V_b$ = 0.46 V. Although much smoothened out compared with *electrochemical* potentials,[33] *electrostatic* potentials also show a corresponding transition from the symmetric low-bias ($V_b \leq 0.3$ V) to asymmetric high-bias ($V_b \geq 0.3$ V) profiles (Supporting Information Figure S13). While we recently showed[34] that this "quantum-hybridization NDR" mechanism is also the origin of the NDR signals recently observed in two-dimensional van der Waals heterostructure transistors,[35-37] the perovskite nanowires with a negligible Schottky barrier or near-Ohmic contacts[16] at the PbI$_3$-TMSPbI$_3$ interfaces apparently produce higher current densities and PVRs.

### 3. Conclusions

In summary, taking the recently synthesized TMSPbI$_3$ hybrid halide perovskite as a representative example, we carried out first-principles calculations to investigate the material properties and electronic device applications of halide perovskite nanowire. The energetic and dynamical stabilities of low-dimensional TMSPbI$_3$ forms were confirmed by computing exfoliation energies and phonon band dispersions, respectively. In the process of nanostructuring from the 3D bulk to the 1D nanowire limits, these structures showed very small variations in their electronic and optical spectra, revealing their intrinsically 1D nature. Notably, the inorganic core of TMSPbI$_3$, the face-sharing octahedral PbI$_3$ framework, also turns out be structurally stable and moreover assume a semimetallic property. Given the possibility of deriving metallic as well as semiconductor nanowire elements from the parental TMSPbI$_3$ perovskite structure, we finally considered the PbI$_3$-TMSPbI$_3$-PbI$_3$ nanowire junction models and obtained very strong NDR characteristics. We emphasized that the results provide a proof of concept for the novel "quantum-hybridization NDR" mechanism that involves the quantum-mechanical hybridization of channel electronic states with their electrode counterparts at low bias-voltages and its abrupt disruption at higher bias voltages. Our work demonstrates the significant potential of low-dimensional hybrid halide perovskites for post-CMOS electronic devices that will enable, e.g, low-power multi-valued logic applications,[36, 37] encouraging the research community to move beyond the photonic applications for which halide perovskites have been mainly considered.

### 4. Computational Method

*DFT calculations*. The PBEsol[17] and HSE[24] DFT calculations were performed with the Vienna Ab-initio Simulation Package.[38] The plane-waves were expanded with a kinetic energy cutoff of 600 eV to obtain basis sets with the self-consistency cycle convergence energy criterion of $10^{-8}$ eV. A *k*-point mesh of 5 x 5 x 8, 1 x 5 x 8, and 1 x 1 x 8 was employed for 3D, 2D, and 1D structures, respectively. Atomic structures were optimized using conjugate-gradient approach until the Hellmann–Feynman forces were less than 0.001 eV/Å. The simulations were performed within the PBEsol generalized gradient approximation.[17] The core and valence electrons were handled by the projector augmented wave method.[38] A vacuum space of more than 15 Å was inserted perpendicular to the periodic directions of 2D nanosheet and 1D nanowire perovskites to avoid interactions with their neighboring images in periodic boundary condition setup. In order to determine the dynamic stability of (CH$_3$)$_3$SPbI$_3$ nanostructures, we adopted the 2 x 2 x 3, 1 x 3 x 3, and 1 x 1 x 4 supercells for of 3D and 2D, and 1D cases, and computed the force constant matrices by using the small displacement method. In the electronic band structure calculations, we included the SOC and its effects are discussed in Supporting Information (Table S2 and Figure S5).

*DFT-NEGF calculations*. Quantum transport calculations were performed by using the `TranSIESTA` software[39] that implements the NEGF method.[33, 40] The PBEsol generalized gradient approximation[17] was employed together with the double-ζ plus polarization-quality atomic orbital basis sets. The device geometry was optimized until atomic forces become smaller than 0.04 eV/Å. The MPH orbitals were generated using the *Inelastica* code.[31]

## Acknowledgements

This work was supported by the Nano-Material Technology Development Program (Nos. 2016M3A7B4024133 and 2016M3A7B4909944), Basic Research Program (No. 2017R1A2B3009872), Global Frontier Program (No. 2013M3A6B1078881), and Basic Research Lab Program (No. 2017R1A4A1015400) of the National Research Foundation funded by the Ministry of Science and ICT of Korea. Computational resources were provided by the KISTI Supercomputing Center (KSC--2017-C3-0085).

## Author Information

M.E.K. and J.L. contributed equally to this work.
*Y.-H.K.: corresponding author; y.h.kim@kaist.ac.kr

## References

[1] Q. A. Akkerman, G. Raino, M. V. Kovalenko, L. Manna, *Nat. Mater.* **2018**, 17, 394.
[2] R. E. Brandt, V. Stevanović, D. S. Ginley, T. Buonassisi, *MRS Commun.* **2015**, 5, 265.
[3] M. V. Kovalenko, L. Protesescu, M. I. Bodnarchuk, *Science* **2017**, 358, 745.
[4] H. J. Snaith, *Nat. Mater.* **2018**, 17, 372.
[5] W.-J. Yin, J.-H. Yang, J. Kang, Y. Yan, S.-H. Wei, *J. Mater. Chem. A* **2015**, 3, 8926.
[6] L. Dou, A. B. Wong, Y. Yu, M. Lai, N. Kornienko, S. W. Eaton, A. Fu, C. G. Bischak, J. Ma, T. Ding, N. S. Ginsberg, L.-W. Wang, A. P. Alivisatos, P. Yang, *Science* **2015**, 349, 1518.
[7] M. Yuan, L. N. Quan, R. Comin, G. Walters, R. Sabatini, O. Voznyy, S. Hoogland, Y. Zhao, E. M. Beauregard, P. Kanjanaboos, Z. Lu, D. H. Kim, E. H. Sargent, *Nat. Nanotechnol.* **2016**, 11, 872.
[8] Z. Yuan, C. Zhou, Y. Tian, Y. Shu, J. Messier, J. C. Wang, L. J. van de Burgt, K. Kountouriotis, Y. Xin, E. Holt, K. Schanze, R. Clark, T. Siegrist, B. Ma, *Nat. Commun.* **2017**, 8, 14051.
[9] Y. Zhang, J. Liu, Z. Wang, Y. Xue, Q. Ou, L. Polavarapu, J. Zheng, X. Qi, Q. Bao, *ChemComm* **2016**, 52, 13637.
[10] H. Lin, C. Zhou, Y. Tian, T. Siegrist, B. Ma, *ACS Energy Lett.* **2018**, 3, 54.
[11] D. B. Mitzi, C. A. Feild, W. T. A. Harrison, A. M. Guloy, *Nature* **1994**, 369, 467.
[12] S. Wang, D. B. Mitzi, C. A. Feild, A. Guloy, *J. Am. Chem. Soc.* **1995**, 117, 5297.
[13] G. A. Mousdis, V. Gionis, G. C. Papavassiliou, C. P. Raptopoulou, A. Terzis, *J. Mater. Chem.* **1998**, 8, 2259.
[14] A. Kaltzoglou, C. C. Stoumpos, A. G. Kontos, G. K. Manolis, K. Papadopoulos, K. G. Papadokostaki, V. Psycharis, C. C. Tang, Y.-K. Jung, A. Walsh, M. G. Kanatzidis, P. Falaras, *Inorg. Chem.* **2017**, 56, 6302.
[15] A. Kaltzoglou, M. M. Elsenety, I. Koutselas, A. G. Kontos, K. Papadopoulos, V. Psycharis, C. P. Raptopoulou, D. Perganti, T. Stergiopoulos, P. Falaras, *Polyhedron* **2018**, 140, 67.
[16] F. Leonard, A. A. Talin, *Nat. Nanotechnol.* **2011**, 6, 773.
[17] J. P. Perdew, A. Ruzsinszky, G. I. Csonka, O. A. Vydrov, G. E. Scuseria, L. A. Constantin, X. Zhou, K. Burke, *Phys. Rev. Lett.* **2008**, 100, 136406.
[18] J. T. Paul, A. K. Singh, Z. Dong, H. Zhuang, B. C. Revard, B. Rijal, M. Ashton, A. Linscheid, M. Blonsky, D. Gluhovic, J. Guo, R. G. Hennig, *J. Phys. Condens. Matter* **2017**, 29, 473001.
[19] S. Grimme, J. Antony, S. Ehrlich, H. Krieg, *J. Chem. Phys.* **2010**, 132, 154104.
[20] H. Häkkinen, *Nat. Chem.* **2012**, 4, 443.
[21] S. S. Jang, Y. H. Jang, Y. H. Kim, W. A. Goddard, 3rd, A. H. Flood, B. W. Laursen, H. R. Tseng, J. F. Stoddart, J. O. Jeppesen, J. W. Choi, D. W. Steuerman, E. Deionno, J. R. Heath, *J. Am. Chem. Soc.* **2005**, 127, 1563.
[22] Y.-H. Kim, S. S. Jang, W. A. Goddard III, *J. Chem. Phys.* **2005**, 122, 244703.
[23] E. W. Elliott, R. D. Glover, J. E. Hutchison, *ACS Nano* **2015**, 9, 3050.
[24] J. Heyd, G. E. Scuseria, M. Ernzerhof, *J. Chem. Phys.* **2006**, 124, 219906.
[25] M. Rohlfing, S. G. Louie, *Phys. Rev. Lett.* **1998**, 81, 2312.
[26] Y.-H. Kim, A. Görling, *Phys. Rev. Lett.* **2002**, 89, 096402.
[27] A. R. Goñi, A. Pinczuk, J. S. Weiner, J. M. Calleja, B. S. Dennis, L. N. Pfeiffer, K. W. West, *Phys. Rev. Lett.* **1991**, 67, 3298.
[28] T. Nagao, S. Yaginuma, T. Inaoka, T. Sakurai, *Phys. Rev. Lett.* **2006**, 97, 116802.
[29] H. A. Atwater, A. Polman, *Nat. Mater.* **2010**, 9, 205.
[30] W. Zhang, M. Saliba, S. D. Stranks, Y. Sun, X. Shi, U. Wiesner, H. J. Snaith, *Nano Lett.* **2013**, 13, 4505.
[31] M. Paulsson, M. Brandbyge, *Phys. Rev. B* **2007**, 76, 115117.
[32] G. I. Lee, J. K. Kang, Y.-H. Kim, *J. Phys. Chem. C* **2008**, 112, 7029.
[33] S. Datta, *Electronic Transport in Mesoscopic Systems*, Cambridge University Press, 1995.
[34] H. S. Kim, Y.-H. Kim, *arXiv:1808.03608 [cond-mat.mes-hall]* **2018**.
[35] L. Britnell, R. V. Gorbachev, A. K. Geim, L. A. Ponomarenko, A. Mishchenko, M. T. Greenaway, T. M. Fromhold, K. S. Novoselov, L. Eaves, *Nat. Commun.* **2013**, 4, 1794.
[36] A. Nourbakhsh, A. Zubair, M. S. Dresselhaus, T. Palacios, *Nano Lett.* **2016**, 16, 1359.
[37] J. Shim, S. Oh, D.-H. Kang, S.-H. Jo, M. H. Ali, W.-Y. Choi, K. Heo, J. Jeon, S. Lee, M. Kim, Y. J. Song, J.-H. Park, *Nat. Commun.* **2016**, 7, 13413.
[38] G. Kresse, D. Joubert, *Phys. Rev. B* **1999**, 59, 1758.
[39] M. Brandbyge, J.-L. Mozos, P. Ordejón, J. Taylor, K. Stokbro, *Phys. Rev. B* **2002**, 65, 165401.
[40] Y.-H. Kim, S. S. Jang, Y. H. Jang, W. A. Goddard III, *Phys. Rev. Lett.* **2005**, 94, 156801.

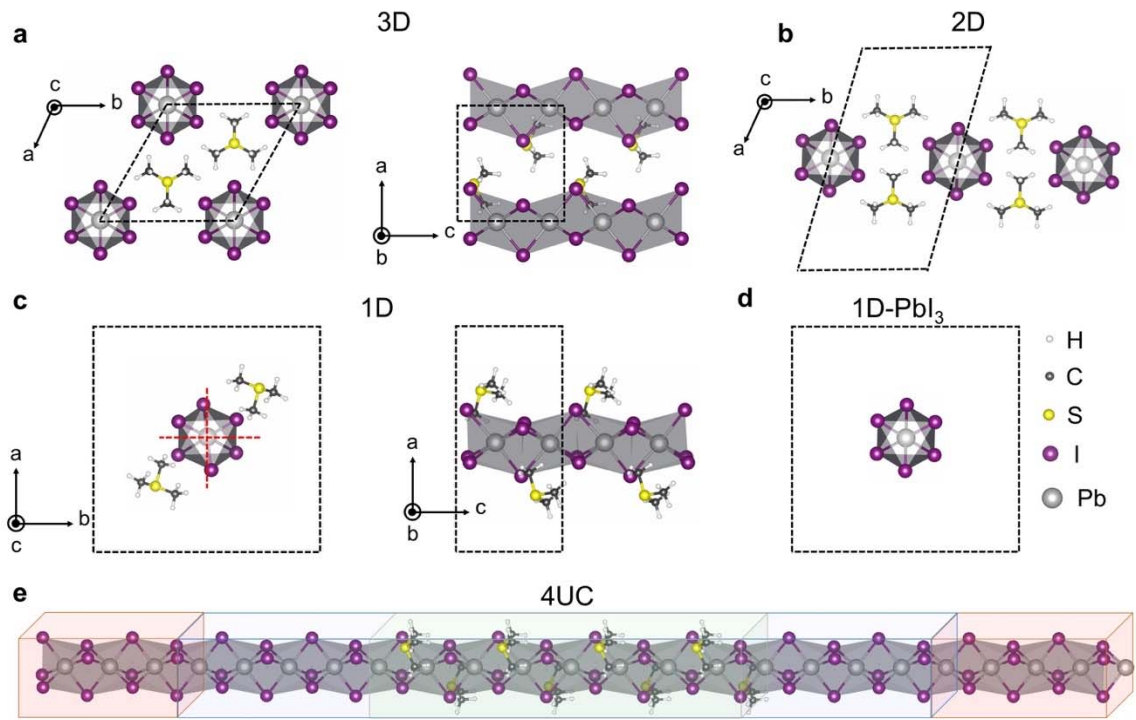

**Figure 1 |** Crystal structures of low-dimensional TMSPbI$_3$ analogs and a NDR device model. The atomic geometries in a) 3D bulk, b) 2D monolayer, c) 1D nanowire, and d) 1D PbI$_3$ inorganic core-only analogs. Distortion or symmetry breaking in the inorganic face-sharing octahedral framework can be noted in the left panel of c. The black dotted boxes in a - d indicate the unit cells for each case. e) The optimized atomic structure of bundled nanowire junctions based on PbI$_3$ electrodes and TMSPbI$_3$ channels (green box region). Red and blue (including green) box regions in e indicate the (semi-infinite) electrodes and central scattering regions, respectively, within NEGF transport calculations.

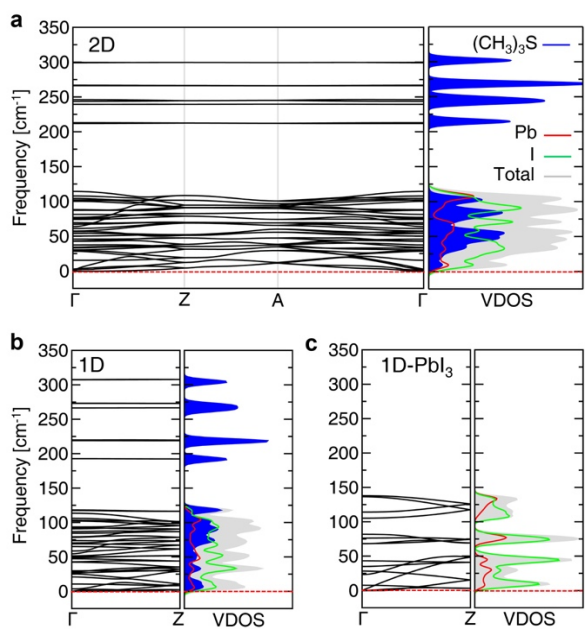

**Figure 2 |** Dynamic stability of TMSPbI3-derived nanostructures. Harmonic phonon band and DOS spectra obtained from a) 2D TMSPbI3, b) 1D TMSPbI3, and c) 1D PbI3 structures calculated within PBEsol. For the DOS, we show the projections to Pb, I, and (CH3)3S. For clarity, high-frequency phonon modes (>350 cm-1) originating from (CH3)3S+ ligands were omitted.

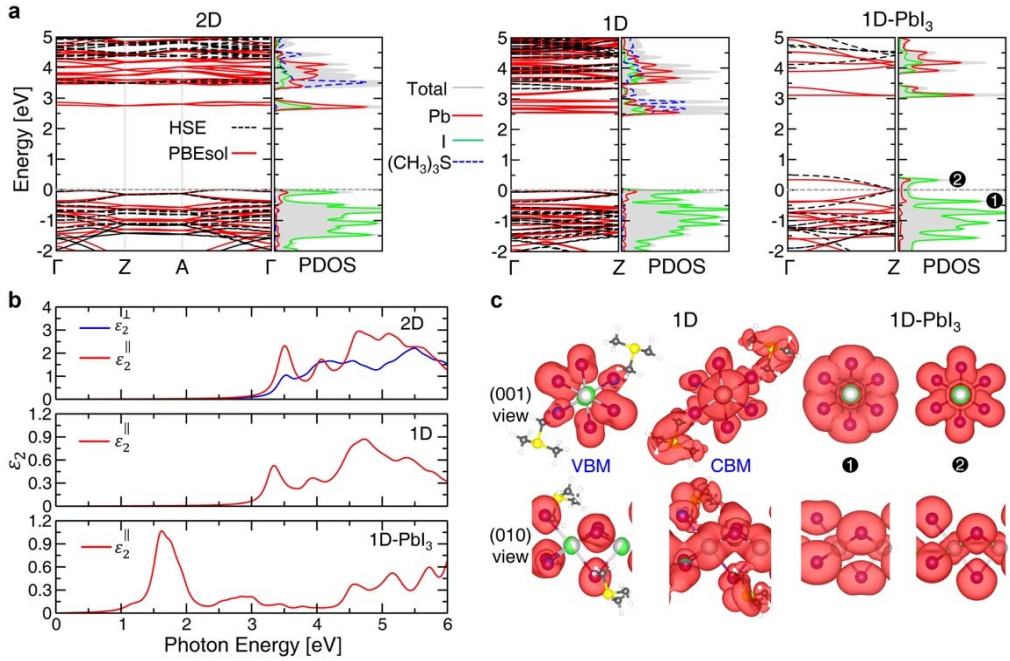

**Figure 3** | Electronic and optical properties low-dimensional forms of TMSPbI$_3$. a) Electronic band structures and projected DOS of 2D and 1D semiconducting TMSPbI$_3$, and 1D semimetallic PbI$_3$ calculated within HSE. The top of occupied bands is set to zero in 2D and 1D TMSPbI$_3$ cases. For the band structures, we also show the PBEsol results. In the DOS plots, we show the projections to Pb, I, and (CH$_3$)$_3$S. b) The imaginary parts of the complex dielectric functions of 2D TMSPbI$_3$, 1D TMSPbI$_3$, and 1D PbI$_3$ calculated within HSE. Blue and red curves in the 2D TMSPbI$_3$ case are the optical spectra along the TMSPbI$_3$ column axial ($\varepsilon_2^{\parallel}$) and normal ($\varepsilon_2^{\perp}$) directions. c) Integrated charge densities around the band edges of 1D TMSPbI$_3$ and PbI$_3$. While VBM and CBM band edge states are shown for 1D TMSPbI$_3$, two VBM peak states are shown for 1D PbI$_3$ indicated as ① and ②. The isosurface level is $3\times 10^{-3}$ e·Å$^{-3}$.

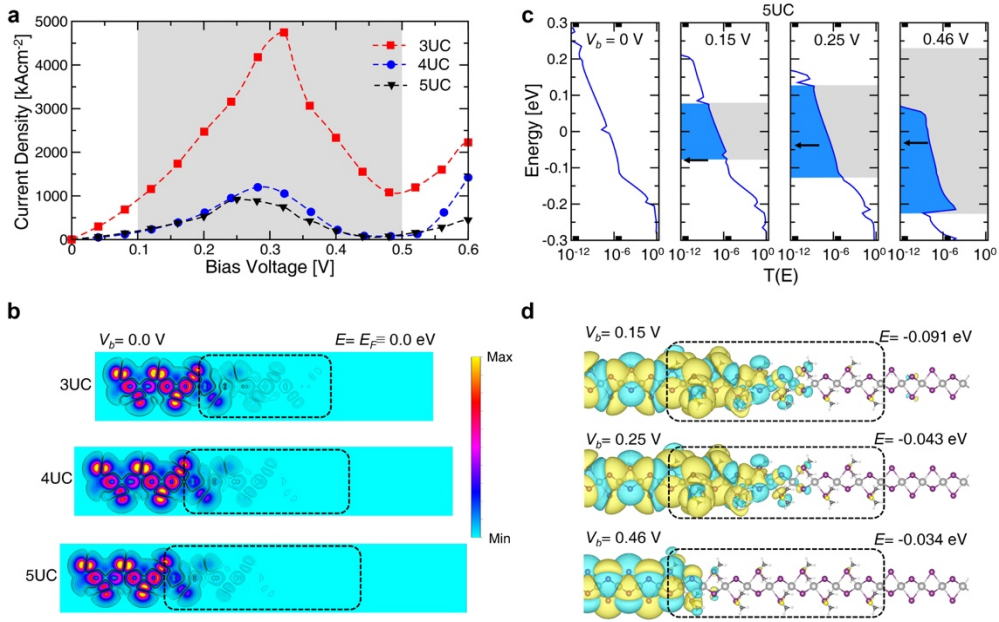

**Figure 4** | The NDR device properties from TMSPbI$_3$ perovskite nanowires. a) The I-V characteristics of the 4UC and 5UC channel junctions. b) Left transmission eigenchannels at V$_b$ = 0.0 V. c) Transmission spectra for 5UC device at different V$_b$ values. Grey shaded regions indicate the bias windows. Note that V$_b$ = 0.25 V and V$_b$ = 0.46 V correspond to the NDR peak and NDR valley regions, respectively. d) Molecular projected Hamiltonian states for the 5UC case at different V$_b$ values that contribute most strongly to the transmissions. The isosurface level is $5\times10^{-4}$ e·Å$^{-3}$.

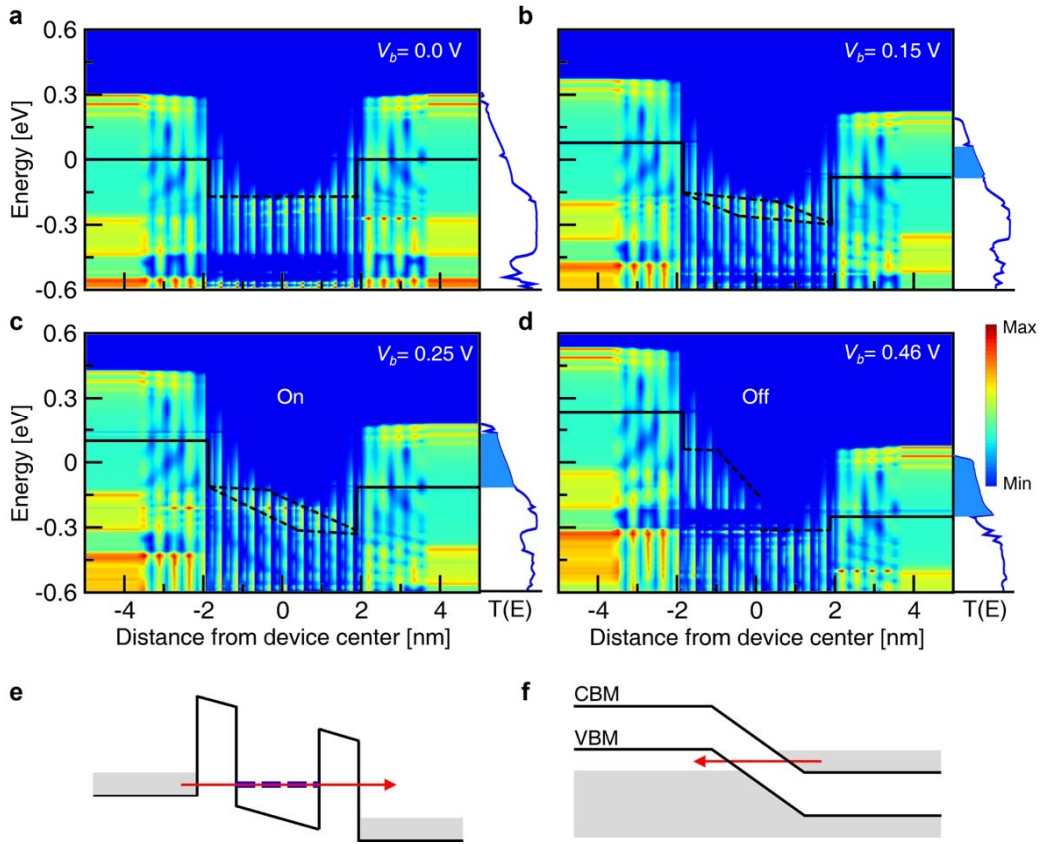

**Figure 5** | The "quantum-hybridization NDR" mechanism. Projected local DOS and energy band diagrams of the 4UC model at a) $V_b = 0.0$ V, b) $V_b = 0.15$ V, c) $V_b = 0.25$ V, and d) $V_b = 0.46$ V. The CBM-region data are not shown for clarity. The solid and dotted lines indicate the Fermi levels in the PbI$_3$ electrodes and the quasi-Fermi levels in the TMSPbI$_3$ channel, respectively. For comparison, the band alignments at the NDR peak regimes in e) the resonant tunneling diode based on the double-barrier tunneling and f) the tunnel diode based on the band-to-band tunneling are schematically depicted.

**Table I** | NDR characteristics of devices based on 3UC, 4UC and 5UC channel models.

| Model | Channel length [nm] | Peak voltage [V] | Valley voltage [V] | Peak current density [kA.cm$^{-2}$] | Valley current density [kA.cm$^{-2}$] | PVR |
|---|---|---|---|---|---|---|
| 3UC | 2.4 | 0.32 | 0.49 | 4753.8 | 1058.5 | 4.5 |
| 4UC | 3.2 | 0.28 | 0.47 | 1195.1 | 73.5 | 16.3 |
| 5UC | 4.0 | 0.25 | 0.46 | 920.7 | 52.8 | 17.4 |